\documentclass[12pt]{article}
\usepackage{sc3conf}
\usepackage{amsfonts}
\usepackage{epsfig}

\newcommand{\JHEP}[3]{JHEP~{\bf #1}{(#2)}{#3}}

\newcommand{\JMP}[3]{J. Math. Phys. {\bf #1}{(#2)}{#3}}
\newcommand{\NP}[3]{Nucl. Phys. {\bf #1}{(#2)}{#3}}
\newcommand{\PL}[3]{Phys. Lett. {\bf #1}{(#2)}{#3}}
\newcommand{\PR}[3]{Phys. Rev. {\bf #1}{(#2)}{#3}}
\newcommand{\PRL}[3]{Phys. Rev. Lett. {\bf #1}{(#2)}{#3}}
\newcommand{\PTP}[3]{Prog. Theo. Phys. {\bf #1}{(#2)}{#3}}

\def\beq{\begin{equation}}
\def\eeq{\end{equation}}
\def\beqn{\begin{eqnarray}}
\def\eeqn{\end{eqnarray}}

\begin{document}
\raggedbottom

\renewcommand{\thefootnote}{\fnsymbol{footnote}}

\title{Off-shell Boundary/Crosscap States and Orientifold Planes$^{*}$}\footnote[0]{$^{*}$Talk given by H.I. at the 3rd Sakharov International Conference on Physics, Moscow, June, 2002.}

\renewcommand{\thefootnote}{\arabic{footnote}}

\authors{H.~Itoyama,\adref{1}  
  and S.~Nakamura\adref{2}}

\addresses{\1ad Department of Mathematics and Physics,
             Graduate School of Science\\
             Osaka City University\\
             3-3-138, Sugimoto, Sumiyoshi-ku, Osaka 558-8585, Japan
\nextaddress \2ad Theoretical Physics Laboratory\\
RIKEN (The Institute of Physical and Chemical Research)\\
2-1 Hirosawa, Wako, Saitama 351-0198, Japan}

\maketitle

\begin{abstract}
The authors' recent works on off-shell boundary/crosscap states are
reviewed.
\end{abstract}

\section{Introduction}

It is well-known that extending string theory off-shell is a hard problem. In the first quantized string theory based on an action
of a single string, the guiding principle is a local Weyl
invariance of string worldsheets. This immediately
translates into on-shellness of string scattering amplitudes.
Off-shellness alone does not suggest any particular symmetry
and one has to consider insertions of operators 
carrying dimensions of arbitrarily high degrees on the bulk
of the string worldsheets: this is a hard
task to carry out. Situation can be more manageable to handle if
the local Weyl invariance is broken  only on the boundaries of
the worldsheets.

To date there are two proposals for off-shell string theory
through deformation on the boundary of the string worldsheets.
The one is the boundary string field theory (BSFT) of Witten 
\cite{wittenbsft}
on a unit disc and the other is  the sigma model approach
of open strings to the
spacetime action $S$. 
(See for example Ref. \cite{tseytlin11/2000} and the references
therein.)
To put simply,  
\begin{equation}
  \label{eq:1}
 Z^{single} \longrightarrow  S \;\;,
\end{equation}
where $Z^{single}$ is a partition function for a single string.
The former proposal was successfully applied to the problem of
open string tachyon condensation
\cite{open-tachyon}.
It is well known that open strings alone do not serve
as perturbatively stable system in flat spacetime
\cite{Tadcan}. 
A possible next step common to these two approaches will be
to extend the idea to the other worldsheet geometry which
shares the same Euler number as that of a disc, namely $RP^{2}$. 
This amounts to considering the string field theory of
unoriented open and closed strings at least at the linearized
level \cite{mixed}.
This was considered and developed in Refs. \cite{Inaka1, Inaka2}.

The above thoughts have also suggested  that the off-shell
extension of boundary states \cite{OBS}
and that of crosscap states \cite{Inaka1, Inaka2}
are an expedient tool and can be used in more general contexts.
In particular, the computation of $g$ function  is equivalent
to the determination of the normalization of these states.
(See Refs. \cite{Fujii-Itoyama-1,Ito-Oota} and the references therein.)

\section{BSFT and off-shell boundary state}

BSFT \cite{wittenbsft}
is defined on a disc worldsheet.
Let us consider an unit disc $\Sigma$ with worldsheet
action
\beqn
I^{disc}
=
\frac{1}{2\pi \alpha'} \int_{\Sigma} d^{2} z
\partial X^{\mu} \bar{\partial}X_{\mu}
+
\int_{\partial \Sigma}
 \frac{d\sigma}{2\pi} {\cal V}\left( \sigma\right),
\label{disc-action}
\eeqn
where ${\cal V}$ is a generic scalar operator 
with ghost number $0$ 
Let us consider the operator
${\cal O}(\sigma)=c^{\sigma}(\sigma){\cal V}(w, \bar{w})$
located on the boundary, where
$c^{\sigma}\left( \sigma\right)$ is the tangent component of 
the ghost field along the boundary.

The defining differential equation
for $S^{disc}$ is written as
\beqn
 \frac{\delta S^{disc} }{\delta \lambda_{\alpha}}
&\propto &
tr\int \int \frac{d\sigma}{2\pi}
    \frac{d\sigma^{\prime}}{2\pi}
\langle {\cal O}^{\alpha}\left( \sigma\right)
\{ Q_{BRS}, {\cal O} \}\left( \sigma^{\prime} \right)
 \rangle_{ {\cal V}, disc}^{ghost} \;\;\;, 
\nonumber \\
{\cal O}\left( \sigma\right) &=&  c^{\sigma}\left( \sigma\right)
{\cal V}\left( \sigma\right)  = \sum_{\alpha}\lambda_{\alpha}
c^{\sigma}\left( \sigma \right) {\cal V}^{\alpha} \left( \sigma\right)
  = \sum_{\alpha}\lambda_{\alpha} {\cal O}^{\alpha}\left( \sigma\right)
\;\;,
\eeqn
where $tr$ denotes the trace over the Chan-Paton space.
The BRS charge $Q_{BRS}$ corresponds to a fermionic
vector field $V$ which is a basic ingredient of the
formalism of BSFT.
Here
$\langle \cdots \rangle_{ {\cal V}, disc}^{ghost}$
is the unnormalized path integral with respect to 
the worldsheet action (\ref{disc-action}),
and can be represented as a matrix element
$\langle \cdots \rangle_{ {\cal V}, disc}=
\langle B \mid \cdots \mid 0 \rangle_{ {\cal V},  disc}$.
The bra vector  $\langle  B \mid$ obeys
\beq
\label{Bcondition}
 \langle  B \mid    \left. \left(  \frac{1}{2\pi \alpha^{\prime}}
 \left( z \frac{\partial}{\partial z} +
\bar{z}
 \frac{\partial}{\partial\bar{z} } \right)  X^{\mu} +   \frac{1}{2 \pi}
 \frac{\partial {\cal V}}{\partial X^{\mu}}  \right)
 \right|_{z=e^{i\sigma}, \; \bar{z}= e^{-i\sigma} }    =  0\;\;.
\eeq
We refer to $\langle B \mid $ as an off-shell 
boundary state (OBS) of the disc.
The condition (\ref{Bcondition}) is a consequence from
the correspondence between the matrix element
and the end point of the path integral. 
The latter one obeys the boundary condition derived 
from the worldsheet action $I^{disc}$ on the disc.
Eq. (\ref{Bcondition}) tells us that, at the initial point 
of the coupling constant flow, the system obeys the
Neumann  boundary condition,
and the end point of the flow is described by zero of
$\frac{\partial {\cal V}}{\partial X^{\mu}} $, namely 
the Dirichlet boundary condition.

\section{Extension of BSFT on $RP^{2}$ Worldsheet Geometry}

We now proceed to construct the $RP^{2}$ extension \cite{Inaka1}
of BSFT.
$RP^{2}$ is a non-orientable Riemann surface of Euler number one 
with no hole, no boundary and one crosscap.
We construct the $RP^{2}$ worldsheet on a complex $z$-plane 
by using an involution where we identify $z$ and $-1/\bar{z}$. 
We choose the fundamental region $\Sigma'$ to be
$\{z=r e^{i\sigma} |0\le r < 1,0\le \sigma <2\pi\}\cup
\{z=r e^{i\sigma} |r= 1,0\le \sigma <\pi\}$.
The crosscap ${\cal C}$, the non-trivial closed loop of the $RP^{2}$
worldsheet, is represented as half of unit circle
$\{z=r e^{i\sigma} |r= 1,0\le \sigma <\pi\}$
in this case.

Let us consider the $RP^{2}$ worldsheet with the following
action:
\beqn
I^{RP^2} =
\frac{1}{2\pi \alpha'}
\int_{\Sigma^{\prime} -{\cal C} } 
d^{2} z 
\partial X^{\mu} \bar{\partial} X_{\mu} 
+
\int_{ {\cal C}}
\frac{d\sigma}{2\pi} {\cal V}^{\prime} \left( \sigma\right)
\;\;.
\label{rp2-action}
\eeqn
Our first objective is to obtain
the operator
which corresponds to the fermionic vector field $V$ 
for the $RP^2$ case.
Let us consider the following operator:
\begin{eqnarray}
Q^{(\sigma)}_{BRS}
&\equiv&
\oint \frac{dz}{2\pi i}
\left( j_z^{(g)}(z)+j_z^{(m)}(z) \right) \;\;,
\nonumber \\
j_z^{(m)}(z)
&\equiv& 2
c^z_{\mbox{\scriptsize even}}(z)
\left( -\frac{1}{\alpha '}\right)
:(\partial_z X^{\mu})_{\mbox{\scriptsize even}}(z)
(\partial_z X_{\mu})_{\mbox{\scriptsize even}}(z): \;\;,
\nonumber \\
j_z^{(g)}(z)
&\equiv&
:b_{zz\ \mbox{\scriptsize even}}(z)
c^z_{\mbox{\scriptsize even}}(z)
\partial_z c^z_{\mbox{\scriptsize even}}(z): \;\;,
\label{maru2}
\end{eqnarray}
where  the subscript $\mbox{\scriptsize even}$ implies that
the modes are restricted to the even ones.

We can show that
$\delta_{B} c^{\sigma}(w)=i\epsilon 
c^{\sigma}\partial_{\sigma} c^{\sigma}(w)$
and $\delta_{B} c^{r}(w)=0$,
where $c^{r}$ is the normal component of the ghost field
along the loop. ($|w|=1$ is understood.)
We also obtain
$\delta_{B}X^{\mu}=i\epsilon c^{\sigma}\partial_{\sigma} X^{\mu}(w)$.
Thus, the operator $Q^{(\sigma)}_{BRS}$
generates the BRS transformations $\delta_{B}$
associated with the diffeomorphisms 
in the $\sigma$ direction on $|w|=1$.

It is now immediate to carry out the action of $Q^{(\sigma)}_{BRS}$
on a generic operator with ghost number one.
This also brings us an operator which is "on-shell" with respect to
$Q^{(\sigma)}_{BRS}$,  that is,
invariant under $\delta_{B}$.
Let ${\cal O}$ be a scalar operator with ghost number one,
${\cal O}(w,\bar{w})=
c^{\sigma}(w){\cal V}^{ \prime \prime}(X^{\mu}(w,\bar{w}))
+c^{r}(w){\cal V}^{ \prime}(X^{\mu}(w,\bar{w}))$,
where ${\cal V}^{\prime}$, 
${\cal V}^{\prime \prime}$ are generic scalar relevant operators.
We can show $\delta_{B}^2 {\cal O}=0$ which is
the condition we need in our formalism.

Following the disc case, we introduce
a defining differential equation 
for $S^{RP^{2}}$
\beqn
\frac{\delta S^{RP^{2}} }{\delta \lambda_{\alpha}}
&\propto &
\int_{\cal C} \int_{\cal C}
 \frac{d\sigma}{2\pi} \frac{d\sigma^{\prime}}{2\pi}
 \langle {\cal O}^{\alpha}\left( \sigma\right)
 \{Q_{BRS}^{(\sigma)} , {\cal O} \}\left( \sigma^{\prime} \right)
 \rangle_{ {\cal V}', RP^{2}}^{ghost} 
\nonumber \\
&=&
2\!\! \int_{\cal C} \!\! \int_{\cal C} \!
 \frac{d\sigma}{2\pi} \frac{d\sigma^{\prime}}{2\pi}
  \sin(\sigma-\sigma')
 \big\langle c_1 c_0 c_{-1} \big\rangle_{RP^{2}} \!\!\
 f(\sigma,\sigma')\;\;, 
\eeqn
where 
$f(\sigma,\sigma')\equiv \bigg\langle
{\cal V}^{\prime \alpha}(\sigma)
\left(
\partial_{\sigma'} {\cal V}^{\prime}(\sigma ')-i \frac{\alpha'}{2}
\frac{\partial^{2}}{\partial X^{\mu} \partial X_{\mu}} 
{\cal V}^{\prime}(\sigma') \right)
\bigg\rangle_{ {\cal V}',  RP^{2}}$,
${\cal V}^{\prime} \equiv  \sum_{\alpha} \lambda_{\alpha}
{\cal V}^{\prime \alpha}$.
Note that the allowed form of
the nonvanishing ghost three point function has selected 
${\cal V}^{\prime}$
alone and  ${\cal V}^{\prime \prime}$ has disappeared.
Here the unnormalized path integral  
$\langle \cdots \rangle_{ {\cal V}'}$ is evaluated 
with respect to the action (\ref{rp2-action}),
and can be represented as a matrix element between 
the ket vector of the closed string vacuum and
the bra vector of the off-shell crosscap state
(OCS) $\langle  C \mid$,
namely 
$\langle \cdots \rangle_{ {\cal V}^{\prime}}=
\langle C \mid \cdots \mid 0 \rangle_{ {\cal V}^{\prime}}$.
Details of the OCS will be given in the next section.

\section{$RP^{2}$ worldsheet with background dilatons} \label{RP2}

In this section, we consider the $RP^{2}$ worldsheet 
with background dilaton field $\Phi$
from the viewpoint of the sigma model approach \cite{Inaka2}.
The basic idea of the sigma model approach is that the spacetime action
for string fields is essentially the renormalized partition 
function of the worldsheet with corresponding background string fields.
In the case of the disc for bosonic strings, for example,
we need corrections of the disc partition function 
in order to subtract the divergence from the M\"{o}bius infinity
\cite{Mobius,tseytlin11/2000}.
However,
the M\"{o}bius group of $RP^{2}$ is $SO(3)$ whose
volume is finite, and we have no 
M\"{o}bius infinity from the $RP^{2}$ worldsheet.
Therefore it is natural to assume that the partition function
of the $RP^{2}$ worldsheet with background fields
itself is the exact loop correction 
term from the $RP^{2}$ graph
in the presence of the quadratic background fields.

To begin with, we set the metric inside the unit circle
($|z| \le 1$) on the complex plane as
$h_{zz}=h_{\bar{z}\bar{z}}=0$,
$h_{\bar{z}z}=h_{z\bar{z}}=1/2$.
The metric outside the unit circle ($|z'| \ge 1$) is obtained
by the involution 
$z'=-\frac{1}{\bar{z}}$, $\bar{z}'=-\frac{1}{z}$
as $h_{z'\bar{z}'}=\frac{1}{r^{4}} h_{\bar{z}z}$,
where $r^{2}=z'\bar{z}'$ for $r \ge 1$.
The worldsheet curvature $R$ is then given by
$\sqrt{g}R=4\left\{
\delta'(r-1) r\ln r +(2+\ln r)\delta(r-1)
\right\}$,
and we obtain
\begin{eqnarray}
\frac{1}{4\pi}
\int_{\Sigma'} dr d\sigma \sqrt{g}R \Phi(r,\sigma)
=
\frac{1}{\pi} \int_{0}^{\pi} d\sigma
\Phi(1,\sigma).
\label{dilaton-RP2}
\end{eqnarray}
Here $g$ is the determinant of the worldsheet metric
written in the polar coordinate $(r,\sigma)$.
Note that Eq. (\ref{dilaton-RP2}) gives the correct Euler number of 
the $RP^2$ (which is one) if we set $\Phi=1$.
Therefore the contribution of the background dilaton concentrates
on the crosscap with the above gauge choice.

\subsection{Off-shell crosscap conditions and OCS}
Let us consider the $RP^{2}$ worldsheet $\Sigma'$ with 
the following action:
\begin{eqnarray}
I=
\frac{1}{2\pi\alpha'}
\int_{\Sigma'-{\cal C}} d^{2} z 
\partial X^{\mu} \bar{\partial}X_{\mu}
+
\frac{1}{\pi}\int_{{\cal C}} d\sigma
\Phi(\sigma),
\label{ws-action}
\end{eqnarray}
where
$\Phi(\sigma)=
a+\frac{1}{2\alpha'}\sum_{\mu=1}^{26}u_{\mu} (X^{\mu}(\sigma))^{2}$.
Note that the worldsheet action is free in the ``bulk" region
$\{z=r e^{i\sigma} |0\le r < 1,0\le \sigma <2\pi\}$
in this gauge choice.
 
The aim of this subsection is to extend the on-shell crosscap conditions
into the case $u_{\mu} \neq 0$.
We should find, in other words, the constraints on 
the closed-string modes in the neighborhood of ${\cal C}$
in the presence of interaction $\Phi$.
We call these constraints off-shell crosscap conditions.
We assert that the off-shell crosscap conditions can be
written as
\begin{eqnarray}
K(z,\bar{z})|_{r \rightarrow 1}
&=& 0,
\nonumber \\
\left\{
(z\partial_z + \bar{z}\bar{\partial}_{\bar{z}})
K(z,\bar{z})
\right\}|_{r \rightarrow 1}
&=& 0,
\label{gen-cross-cond}
\end{eqnarray}
where
\begin{eqnarray}
K(z,\bar{z})
&\equiv&
\left.
\left\{
(w\partial_w + \bar{w}\bar{\partial}_{\bar{w}})
X^{\mu}(w,\bar{w})
+
u_{\mu}X^{\mu}(w,\bar{w})
\right\}
\right|_{w=z,\bar{w}=\bar{z}}
\nonumber \\
&+&
\left.
\left\{
(w\partial_w + \bar{w}\bar{\partial}_{\bar{w}})
X^{\mu}(w,\bar{w})
+
u_{\mu}X^{\mu}(w,\bar{w})
\right\}
\right|_{w=-1/\bar{z},\bar{w}=-1/z}.
\label{K}
\end{eqnarray}
The right-hand side of Eq. (\ref{K}) indicates the meaning of
the off-shell crosscap conditions;
Eq. (\ref{gen-cross-cond}) are the conditions so that
the $X^{\mu}$ in the neighborhood of ${\cal C}$, as well as 
its image by the involution, connects smoothly with the 
$X^{\mu}$ on ${\cal C}$
which obeys the equation of motion
$\left.
\left\{(z\partial_z + \bar{z}\bar{\partial}_{\bar{z}}
+u_{\mu})X^{\mu}(z,\bar{z})
\right\} \right|_{\cal C}=0$.\footnote{
Although $RP^2$ has no boundary,
$(z\partial_z + \bar{z}\bar{\partial}_{\bar{z}})X^{\mu}(z,\bar{z})$
which comes from the total derivative survives only on the
crosscap due to the involution.
}

The off-shell crosscap conditions (\ref{gen-cross-cond})
are rewritten in terms of closed-string modes as
\begin{eqnarray}
K_{1} \equiv
-\{
\alpha_n^{\mu}+(-1)^n \tilde{\alpha}_{-n}^{\mu}
\}
+
\frac{u_{\mu}}{n}
\{
\alpha_n^{\mu}-(-1)^n \tilde{\alpha}_{-n}^{\mu}
\}
&=&0,
\nonumber \\
K_{2} \equiv
-i\alpha' p^{\mu}+u_{\mu}X_0^{\mu}
&=&0,
\label{offshell-mode}
\end{eqnarray}
where we do not sum over $\mu$.
We can easily check that the off-shell crosscap conditions
interpolate between the usual on-shell crosscap conditions 
and their T-duals.

We define OCS
$\langle C(\mbox{\boldmath$u$})|$ 
as
$\langle C(\mbox{\boldmath$u$})|K_{1}=0$,
$\langle C(\mbox{\boldmath$u$})|K_{2}=0$.
The explicit form of $\langle C(\mbox{\boldmath$u$})|$ is
given as
\begin{eqnarray}
\langle C(\mbox{\boldmath$u$})| 
= 
\langle 0|
\exp \left(-\frac{1}{2}X_0^{\mu} A_{\mu \nu} X_0^{\nu} \right)
\exp \left(\sum_{m=1}^{\infty} \tilde{\alpha}_m^{\mu}
    C_{\mu \nu}^{(m)} \alpha_m^{\nu} \right),
\label{boundary-state}
\end{eqnarray}
where
$A_{\mu \nu}
=\frac{1}{\alpha'}u_{\mu} \delta_{\mu \nu}$,
and
$C_{\mu \nu}^{(m)}
=
-\frac{(-1)^m}{m}\frac{m-u_{\mu}}{m+u_{\mu}}\delta_{\mu \nu}$.
We can easily check that this OCS becomes (the
T-dual of) the usual on-shell crosscap state if we take the limit
$u^{\mu} \rightarrow 0$ ($u^{\mu} \rightarrow \infty$).
Therefore the OCS naturally interpolates between
the crosscap state for a higher-dimensional O-plane and that for a
lower-dimensional O-plane.
The coupling constant $u^{\mu}$, which is a parameter of the 
configuration of the background dilaton field, controls 
the dimension of the corresponding O-plane.

\subsection{OCS and  partition function}

Next, we show that the OCS is a useful tool to
evaluate the quantities on the $RP^{2}$ worldsheet.
For example, we can calculate the Green's function and the partition
function on the $RP^{2}$ worldsheet in the presence of interaction
$\Phi(\sigma)$ on the crosscap.
Let us consider one-dimensional target space and omit the superscript
$\mu$ of $X$ and $u$ for simplicity.
The Green's function for this case is given by
\begin{eqnarray}
G(z,w)\!\!\!\!&=&
\frac{\langle C(u)| X(z,\bar{z}) X(w,\bar{w})|0\rangle}{\langle
C(u)|0\rangle}
\nonumber \\
&=&\!\!\!\!
-\frac{\alpha'}{2}
\ln(|z-w|^{2} |1+z\bar{w}|^{2})
+\frac{\alpha'}{u}
-\alpha' u \sum_{k=1}^{\infty}
\frac{(-z\bar{w})^{k}+(-\bar{z}w)^{k}}{k(k+u)}.
\mbox{           }
\label{green-rp2}
\end{eqnarray}
The expectation value of
composite operator $X^{2}(\sigma)$ which is defined as 
\begin{eqnarray}
X^{2}(\sigma)
\equiv
\lim_{\epsilon \rightarrow 0}
\left[
X(\sigma)X(\sigma+\epsilon)-
\left\{
-\frac{\alpha'}{2}\ln|1-e^{i\epsilon}|^{2}
+ (\mbox{const.})
\right\}
\right] ,
\label{def-X2}
\end{eqnarray}
is obtained by substituting $z=e^{i\sigma}$ and 
$w=e^{i(\sigma+\epsilon)}$ into Eq. (\ref{green-rp2}):
\begin{eqnarray}
\langle X^{2}(\sigma) \rangle
=
-\alpha' \ln(2q) -\frac{\alpha'}{u}
-2\alpha' \left[ \Psi\left(\frac{u}{2}\right)-\Psi(u) \right],
\label{X2}
\end{eqnarray}
where
$\Psi(u) \equiv \frac{\frac{d}{du}\Gamma(u)}{\Gamma(u)}$
and we have written the constant in Eq. (\ref{def-X2}) as 
$\alpha' \ln q$ by using a positive constant $q$.
The value for $q$ is ambiguous at this stage and depends on the
renormalization scheme.
We will determine the value for $q$ later.

We can calculate the partition function of the $RP^{2}$
worldsheet by using the relationship
$\frac{d}{du}\ln Z(u)=-\frac{1}{2\alpha'}\langle X^{2}(\sigma) \rangle$,
and then we obtain
\begin{eqnarray}
Z(u)
=
\frac{{\sqrt{2q}}^u \sqrt{u} {\Gamma(\frac{u}{2})}^{2}}{\Gamma(u)}
,
\label{Z(u)}
\end{eqnarray}
up to the overall normalization factor.
In general, the partition function for 26-dimensional target space with
$\Phi(\sigma)=
a+\frac{1}{2\alpha'}\sum_{\mu=1}^{26}u_{\mu} (X^{\mu}(\sigma))^{2}$
on the crosscap can be written as
\begin{eqnarray}
Z(a,\mbox{\boldmath$u$})
\equiv
e^{-a} \prod_{\mu=1}^{26} Z(u_{\mu})
=
e^{-a} \prod_{\mu=1}^{26}
{\left(
  \frac{{\sqrt{2q}}^{u_{\mu}}
  \sqrt{u_{\mu}} {\Gamma(\frac{u_{\mu}}{2})}^{2}}{\Gamma(u_{\mu})}
\right)},
\label{general-Z(u)}
\end{eqnarray}
up to the overall normalization factor.

\section{Applications of OBS and OCS}

\subsection{$g$-function}

OBS's are useful tools for analyses
of two-dimensional systems with boundaries.
In the case the boundary interaction is quadratic, 
the proper form of the Green's function and the $g$-function
has been given by using OBS 
\cite{OBS,Ito-Oota}. 
In the determination of the $g$-function 
which is defined as $\langle 0 \mid B \rangle $,
the overall normalization of the OBS is crucial.
The normalization can be fixed by computing the
partition function of the cylinder worldsheet,
$Z^{cylinder}_{B,B'}$, and demanding
$Z^{cylinder}_{B,B'}=
\langle B'\mid e^{-lH}\mid B \rangle$,
without taking the long cylinder limit \cite{Ito-Oota}.
Here $H$ is the Hamiltonian on the cylinder of length $l$,
and $B$ and $B'$ indicate the boundaries.
The left-hand side is calculated via path integral
with $\zeta$-function regularization.
For example, the $g$-function for bosonic theory
(\ref{disc-action})
with the quadratic boundary interaction 
${\cal V}(\sigma)=u X^{2}$ is given by
\begin{eqnarray}
g(u)=(\alpha'/2)^{1/4}(2\pi u)^{-1/2}\Gamma(u+1) (e/u)^{u}.
\end{eqnarray}
For supersymmetric extension with boundary mass,
the fermionic sector gives
\begin{eqnarray}
g_{+}
=
\frac{\sqrt{2\pi}}{\Gamma(u+1/2)}
\left(\frac{u}{e}\right)^{u}
\:\:\mbox{(for NS sector)},
\:\:\:\:
g_{-}
=\frac{2^{1/4}\sqrt{2\pi u}}{\Gamma(u+1)}
\:\:\mbox{(for R sector)},
\end{eqnarray}
which agree with the results of Refs. \cite{Chat}.

\subsection{Descent relation among O-plane tensions}

The descent relation among O-plane tensions can be also
derived by using the result of Sec. \ref{RP2} \cite{Inaka2}.

Let us define quantity $S_p$ as 
$Z(\mbox{a,\boldmath$u$}) \rightarrow S_p$,
where the limit is taken as
$u^{1},\cdots, u^{p+1} \to 0$ and
$u^{p+2},\cdots, u^{26} \to \infty$.
We do not touch the parameter $a$ here.
According to the argument in Sec. \ref{RP2},
$S_p$ is equal to $V_p \times T_p$ where
$V_p$ and $T_p$ are the volume and the tension of an O$p$-plane.
Here, the dimension of the O-plane is $p+1$ and is defined as 
the number of parameters $u^{\mu}$ which are taken to zero.
Then we can write
\begin{eqnarray}
\frac{S_{25}}{S_{24}}
=
\frac{Z(u_{25})|_{u_{25} \to 0}}{Z(u_{25})|_{u_{25} \to \infty}}
=
\frac{\int dx^{25} V_{24} T_{25}}{V_{24} T_{24}}
=
\frac{\int dx^{25} T_{25}}{T_{24}}.
\end{eqnarray}
We find
\begin{eqnarray}
\lim_{u_{25} \rightarrow 0}Z(u_{25})
=
\int dx^{25} \frac{4}{\sqrt{2\pi\alpha'}},
\:\:\:\:
\lim_{u_{25} \rightarrow \infty}Z(u_{25})
=4\sqrt{\frac{\pi}{2}},
\end{eqnarray}
where $x^{25}$ is the zero mode of $X^{25}$ and
we have correctly extracted the integral 
of the zero-mode part of
$Z(u_{25})$ 
\cite{zero-mode}.
Here we have assigned the value 2 to $q$ because 
we can obtain a finite and non-zero value of $Z(u)$ in the limit
$u\rightarrow \infty$ if and only if $q=2$.
In other words, we have chosen the renormalization scheme in 
Eq. (\ref{def-X2}) so that we obtain a finite and non-zero value 
of $Z(u)$ in the limit $u\rightarrow \infty$.
We therefore obtain
\begin{eqnarray}
\frac{T_{24}}{T_{25}}=\frac{\sqrt{2\pi\alpha'}}{4} 4\sqrt{\frac{\pi}{2}}
=\pi \sqrt{\alpha'}.
\end{eqnarray}
This is precisely the ratio of the tension of an O$24$-plane and
that of an O$25$-plane.
In general, we can show in a similar manner that
$T_{p}/T_{q}=\left( \pi \sqrt{\alpha'} \right)^{q-p}$.

\section{Conclusions}

Off-shell crosscap states as well as off-shell boundary states
are useful tools both for formulating off-shell theory
(worldsheet string field theory or off-shell non-linear sigma
model)
of unoriented open and closed strings,
and for performing computation.

In configuration space of strings,
interpolations of O-planes of various dimensions
are certainly possible as are in D-branes, and
the dilaton background is responsible for this.
One important physical issue is how these interpolations are
materialized as physical processes
in systems with moderate instabilities.

\end{document}